\documentclass[a4paper,11pt]{amsart}

\usepackage{a4wide}
\usepackage{amssymb}
\usepackage{color}
\usepackage{epsf}
\usepackage{graphicx}
\usepackage{bbm}

\newtheorem{theorem}{Theorem}[section]

\theoremstyle{definition}
\newtheorem{defi}[theorem]{Definition}
\newtheorem{ex}[theorem]{Example}
\newtheorem{rem}[theorem]{Remark}

\numberwithin{theorem}{section}

\newcommand{\Z}{\mathbbm{Z}}
\newcommand{\N}{\mathbbm{N}}
\newcommand{\Q}{\mathbbm{Q}}
\newcommand{\R}{\mathbbm{R}}

\newdimen{\standardlabelwidth}
\standardlabelwidth-\standardlabelwidth
  \advance\standardlabelwidth by 2\normalparindent
\newcommand{\standardlabel}[1]{#1\kern\standardlabelwidth}
\begin{document}

%% Dies ist am Ende auszukommentieren.
%\date{\today}q

\title[Discrete tomography of F-type icosahedral model sets]{Discrete tomography of F-type icosahedral model sets}
\author{Christian Huck}
\address{Department of Mathematics and Statistics\\
  The Open University\\ Walton Hall\\ Milton Keynes\\ MK7 6AA\\
   United Kingdom}
\email{c.huck@open.ac.uk}
\thanks{The author was supported by EPSRC via Grant EP/D058465/1.}

\begin{abstract}
We address the problem of uniquely reconstructing F-type icosahedral 
quasicrystals from few images produced by quantitative high
resolution transmission electron microscopy and explain recent results
in the discrete tomography of these sets.
\end{abstract}

\maketitle

\section{Introduction}\label{sec1}

Motivated by the request of materials science for the unique
reconstruction of quasicrystalline structures from a small number of
images produced by quantitative high
  resolution transmission electron microscopy (HRTEM), the {\em discrete tomography}\/ (DT) of F-type icosahedral
  model sets is concerned with the inverse problem of uniquely
reconstructing finite patches of these {\em aperiodic}\/ point sets from their 
({\em discrete parallel}\/) {\em $X$-rays}\/ in several directions. As for the discrete analogue of parallel $X$-rays in
 computerized tomography (CT), the $X$-ray of a finite
subset of Euclidean $3$-space in a certain
direction assigns to each line parallel to this
direction the number of points of the set on this line. In fact, for
some crystals the technique QUANTITEM (\textbf{QU}antitative \textbf{AN}alysis of \textbf{T}he \textbf{I}nformation from \textbf{T}ransmission
\textbf{E}lectron \textbf{M}icroscopy) can approximately measure the number of atoms lying on
lines parallel to directions that guarantee
HRTEM images of high resolution, i.e., yield dense lines in the
crystal; cf. ~\cite{ks,sk}. Therefore, the classical
setting of DT deals with the corresponding inverse problem for (periodic) 
 lattices; cf.~\cite{HK,GG,GGP,Gr}. Since F-type icosahedral model sets are commonly regarded as
good mathematical models for many icosahedral quasicrystals~\cite{HG} and since it is reasonable to expect
that future developments in technology will extend the technique QUANTITEM to
classes of quasicrystals, we feel that it is about time to 
explain recent results in the DT of these sets in a manner that is 
easily accessible for practitioners. In particular, special emphasis
is put on illustrating the problems that arise in the DT of aperiodic
model sets. For detailed proofs 
and related results, we refer the interested reader to~\cite{BG2,BH,H1,H2}.

\section{F-type icosahedral model sets}\label{sec2}

We denote the
golden ratio by $\tau$, i.e., $\tau=(1+\sqrt{5})/2$. Note that $\tau$
is a root of $X^2-X-1\in \Z[X]$, whence it is an
algebraic integer of degree $2$ over
$\Q$. The unique
non-trivial Galois automorphism of the real quadratic number field
$\Q(\tau)=\Q(\sqrt{5})=\Q\oplus\Q\tau$, determined by $\sqrt{5}\mapsto
-\sqrt{5}$, will be denoted by $.'$, hence $\tau'=1-\tau$. Throughout
this text, elements of Euclidean $3$-space will be written as row vectors. The standard face-centred icosahedral module of
quasicrystallography is given by 
$$
\mathcal{M}_{\text{F}}\,\,:=\,\,\Z[\tau](2,0,0)\oplus\Z[\tau](\tau+1,\tau,1)\oplus\Z[\tau](0,0,2)\,,
$$
where $
\Z[\tau]=\Z\oplus\Z\tau$ is the ring of integers in $\Q(\tau)$;
cf.~\cite{baake,BPR}. Clearly, $\mathcal{M}_{\text{F}}$ is a free
$\Z[\tau]$-module of rank $3$ that spans all of $\R^3$. In particular,
it is a free $\Z$-module of rank
$6$. Moreover, $\mathcal{M}_{\text{F}}$ has full {\em icosahedral symmetry}, i.e.,
it is invariant under the action of the rotation group $Y$ of the regular icosahedron centred at the origin $0\in\R^3$ with orientation such that each
coordinate axis passes through the mid-point of an edge, thus
coinciding with $2$-fold axes of the icosahedron. Due to the
connection with the {\em icosian ring}, we prefer to use the scaled
version $L:=\tfrac{1}{2}\mathcal{M}_{\text{F}}\subset\Q(\tau)^3$ instead of $\mathcal{M}_{\text{F}}$ itself;
  cf.~\cite{BPR} and references therein.

\begin{figure}
\centerline{\epsfysize=0.58\textwidth\epsfbox{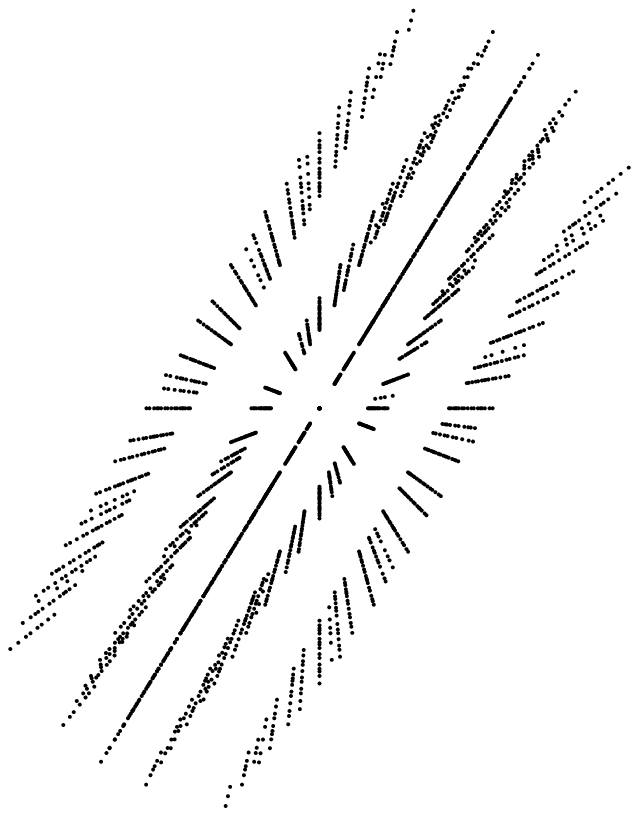}\hspace{0.00\textwidth}
\epsfysize=0.58\textwidth\epsfbox{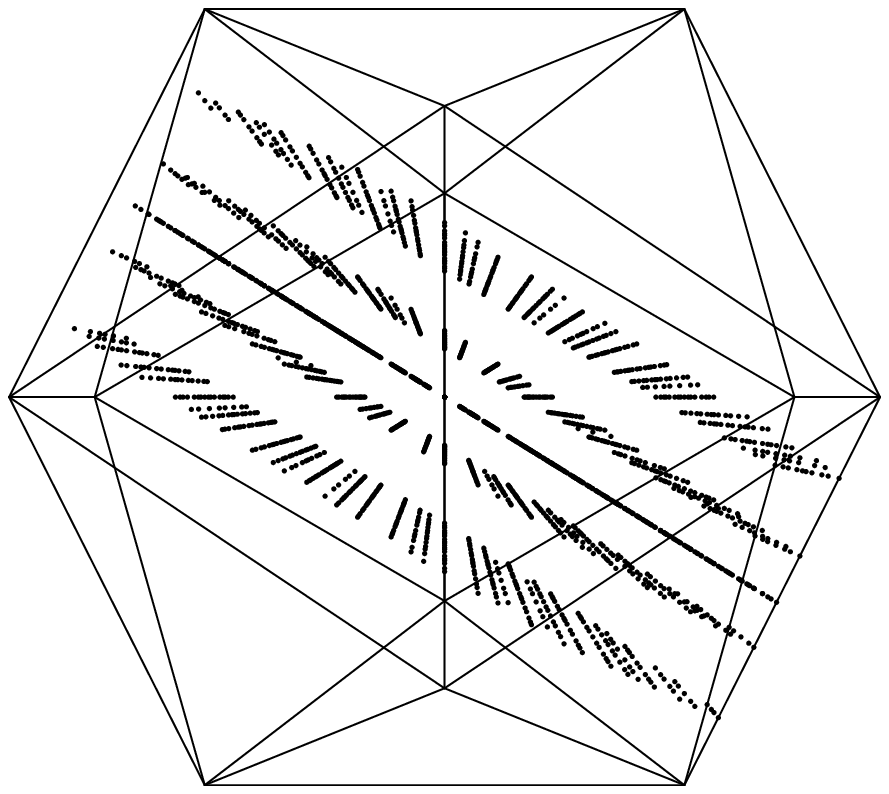}}
\caption{A few slices orthogonal to $(\tau,0,1)$ of a patch of the icosahedral model set
  $\varLambda$ from Example~\ref{lambdaico} (left) and their $.^{\star}$-images
  inside the icosahedral window $s+W$ (right), both seen from the positive $x$-axis.}
\label{fig:slices}
\end{figure}

\begin{defi}\label{modelset}
Let the map $
.^{\star}\!:\, L
\rightarrow \R^3
$
be defined by applying the Galois conjugation $.'$ coordinatewise. Given a subset $W\subset \R^3$ with
$\varnothing\, \neq\, W^{\circ}\subset W\subset \overline{W^{\circ}}$
and $\overline{W^{\circ}}$ compact, and any $t\in\R^3$, we obtain an {\em F-type icosahedral
  model set}\/ 
$\varLambda(t,W)$ by setting 
$$\varLambda(t,W) := t+\{\alpha\in L\,|\,\alpha^{\star}\in W\}$$
The map
$.^{\star}\! : \,  L\rightarrow \R^3$ is the
so-called \emph{star map}\/ of $\varLambda(t,W)$ and $W$ is referred to
as the {\em window}\/ of $\varLambda(t,W)$. The model set
$\varLambda(t,W)$ is called {\em generic}\/ if it satisfies $\partial W\cap
L^{\star}=\varnothing$. Moreover, it is called {\em regular}\/ if the boundary $\partial W$ has Lebesgue
measure $0$ in $\R^3$. For a window $W\subset
\R^3$, we denote by $\mathcal{I}^{\rm F}_{g}(W)$ the set of generic
F-type icosahedral model sets with a window of the form $s + W$, where
$s\in\R^3$.
\end{defi}

We refer the reader to~\cite{H2} for the corresponding cut and project scheme and to~\cite{Moody,BM} for related
general settings and background. F-type icosahedral model sets
$\varLambda$ are {\em aperiodic}\/ Delone subsets of
$3$-space. Moreover, if
$\varLambda$ is regular, then $\varLambda$ is {\em pure point
diffractive}. If
$\varLambda$ is both generic and regular, and, if a suitable translate
of its window has the full icosahedral symmetry of $L^{\star}$, then $\varLambda$ 
 has full {\em icosahedral symmetry}\/ in the sense of symmetries of
 LI-classes,
meaning that a discrete structure has a certain symmetry if the
original and the transformed structure are locally indistinguishable;
cf.~\cite{B,H2} and references therein for details.

\begin{ex}\label{lambdaico}
For a generic regular F-type icosahedral model set with full icosahedral
symmetry, consider $\varLambda:=\varLambda(0,s+W)$, where $s:=10^{-3}(1,1,1)$ and $W$ is the
regular icosahedron centred at the origin $0\in\R^3$ with orientation
such that $(\tau',0,1)$ and $(-\tau',0,1)$ belong to its vertices; see Figure~\ref{fig:slices} for an illustration.     
\end{ex}

\begin{figure}
\centerline{\epsfysize=0.45\textwidth\epsfbox{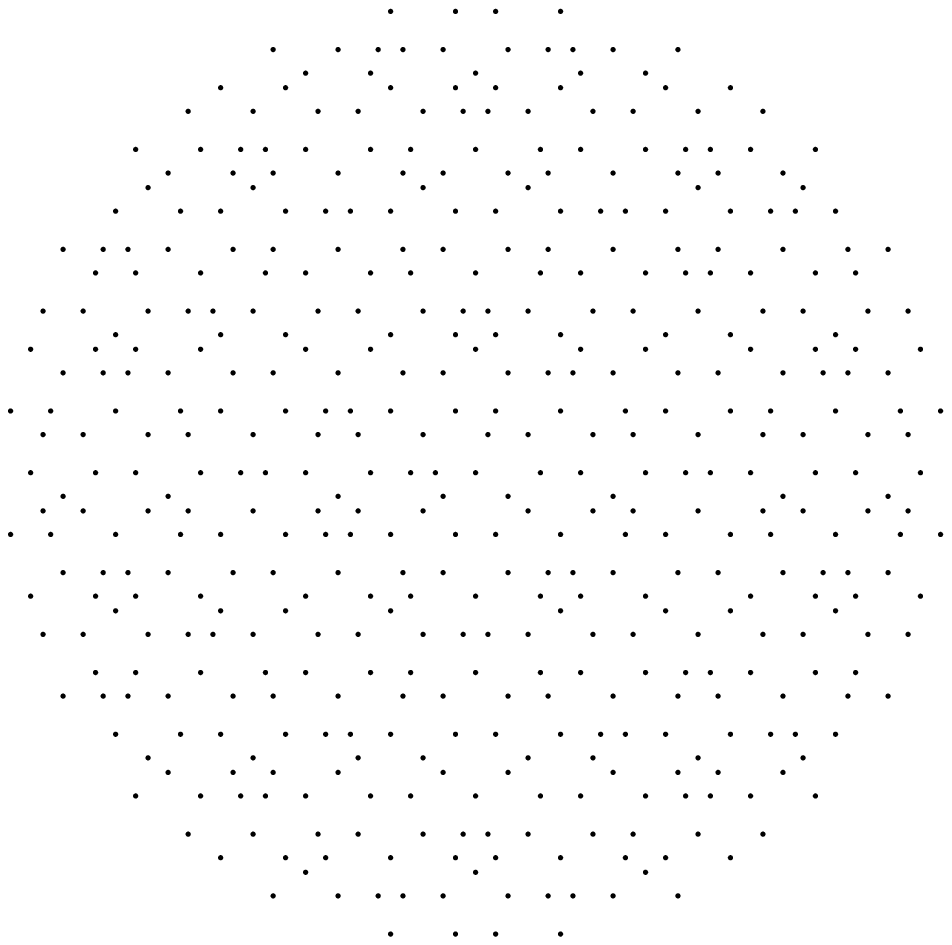}\hspace{0.1\textwidth}
\epsfysize=0.45\textwidth\epsfbox{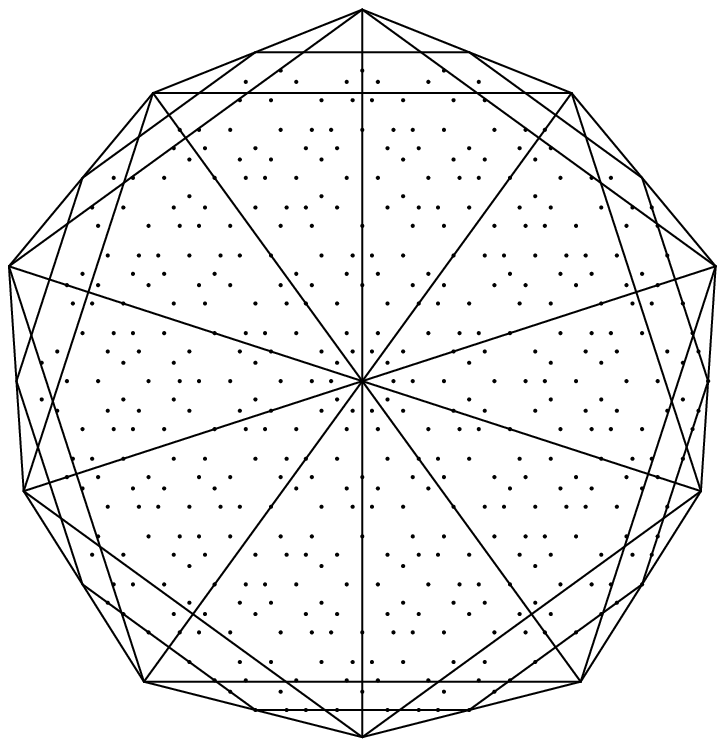}}
\caption{The central slice $S$ of the patch of $\varLambda$ from Figure~\ref{fig:slices} (left) and its
  $.^{\star}$-image $S^{\star}$ inside the (marked) decagon $(s+W)\cap (\tau',0,1)^{\perp}$ (right), both seen from perpendicular viewpoints.}
\label{fig:slices2}
\end{figure}

\section{Complexity}\label{sec3}

The fundamental notion in DT is
the following.

\begin{defi}\label{xray..}
Let $F$ be a finite subset of $\mathbbm{R}^{3}$, let $u\in
\mathbb{S}^{2}$ be a direction, and let $\mathcal{L}_{u}$ be the set
of lines in direction $u$ in $\mathbbm{R}^{3}$. The ({\em
  discrete parallel}\/) {\em X-ray of}\/ $F$ {\em in direction}\/ $u$ is
the function $X_{u}F: \mathcal{L}_{u} \rightarrow
\mathbbm{N}_{0}:=\mathbbm{N} \cup\{0\}$, defined by $$X_{u}F(\ell) :=
\operatorname{card}(F \cap \ell\,) =\sum_{x\in \ell}
\mathbbm{1}_{F}(x)\,.$$ Moreover, the {\em support}\/ $(X_{u}F)^{-1}(\N)$ of $X_{u}F$, i.e.,
the (finite) set of lines in $\mathcal{L}_{u}$ which pass through at least one
point of $F$, is denoted by $\operatorname{supp}(X_{u}F)$. Further, we denote
by $\mathcal{L}_{u}^{L}$ the subset of $\mathcal{L}_{u}$ consisting of
all lines in $\mathcal{L}_{u}$ which pass through at
least one point of $L$.
\end{defi}

\begin{figure}
\centerline{\epsfysize=0.45\textwidth\epsfbox{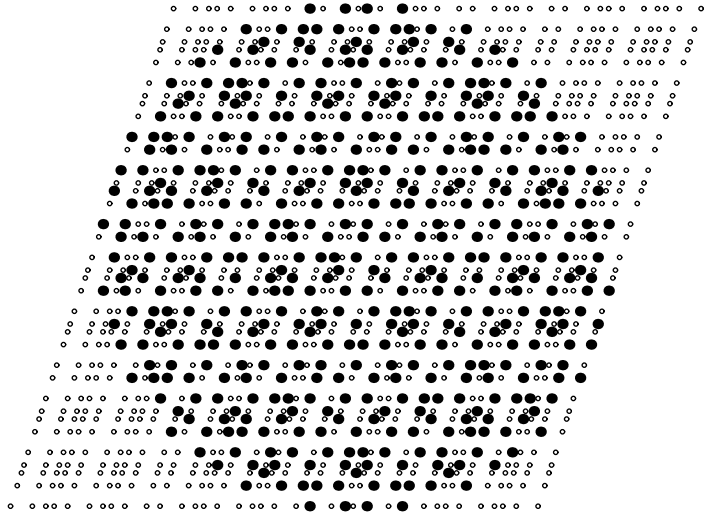}}
\caption{The grid $G$ that is generated by the slice $S$ from
  Figure~\ref{fig:slices2} with respect to two
  $L^{(\tau,0,1)}$-directions with slopes $0$ and
  $\operatorname{tan}(2\pi/5)$, respectively.}
\label{fig:grid1}
\end{figure}

Let $U\subset\mathbb{S}^{2}$
be a finite set of pairwise
non-parallel directions and let $F$ be a finite subset of $\R^3$. Clearly,
$F$ is contained in its {\em grid with respect to the $X$-rays in the directions of $U$} given by
$$
G^{F}_{U}:=\bigcap_{u\in U}\,\, \Big(\bigcup_{\ell \in
  \mathrm{supp}(X_{u}F)} \ell\Big)\,.
$$
In fact, if there are more directions in $U$ than elements in $F$, one
even has 
$F=G^{F}_{U}$; cf.~\cite[Proposition 5.3]{H2}. Since after about $3$ to $5$ images taken by
HRTEM, typical (quasi)crystalline probes may be damaged or even destroyed by the radiation
energy, this result is meaningless in practice. Consequently, one has
to deal with grids that are much bigger than their generating finite
sets; see Figure~\ref{fig:grid1} for an illustration in the plane. In fact, for a finite subset $F$ of a fixed F-type icosahedral model
set, its grid can contain points of a different translate of
$L$ than $F$ itself; see~\cite[Figure 5]{BG2} for illustrations of
this phenomenon in the case of planar sets. As mentioned in the
introduction, only directions that yield dense lines in a fixed F-type
icosahedral model set $\varLambda$ are reasonable in view of
applications in practice. Therefore, we may restrict
ourselves to {\em $\varLambda$-directions}, i.e., directions parallel
to non-zero interpoint vectors of $\varLambda$. It turns out that the set of $\varLambda$-directions
is exactly the set of {\em $L$-directions}\/ (defined analogously) for {\em
  any}\/ 
F-type icosahedral model set $\varLambda$; cf.~\cite[Proposition
  3.20]{H2}.

\begin{defi}[Reconstruction Problem]\label{m-def:consrecprob}
Let $W\subset \mathbbm{R}^3$ be a window and let
  $u_1,\dots,u_m\in\mathbb{S}^2$ be $m\geq 2$ pairwise non-parallel
  $L$-directions. The corresponding reconstruction problem is defined
  as follows.

\flushleft{{\sc Reconstruction}.} \\
  Given functions $p_{u_{j}} : \mathcal{L}_{u_{j}} \rightarrow
  \mathbbm{N}_{0}$, $j\in\{1,\dots,m\}$, whose supports are finite and
  satisfy $\operatorname{supp}(p_{u_{j}})\subset
  \mathcal{L}^{L}_{u_{j}}$, decide whether there exists
  a finite subset $F$ of an element of 
  $\mathcal{I}^{\rm F}_g(W)$ that satisfies
  $X_{u_{j}}F=p_{u_{j}}$, $j\in\{1,\dots,m\}$, and, if so, construct
  one such $F$.

\end{defi}

\begin{figure}
\centerline{
\epsfysize=0.33\textwidth\epsfbox{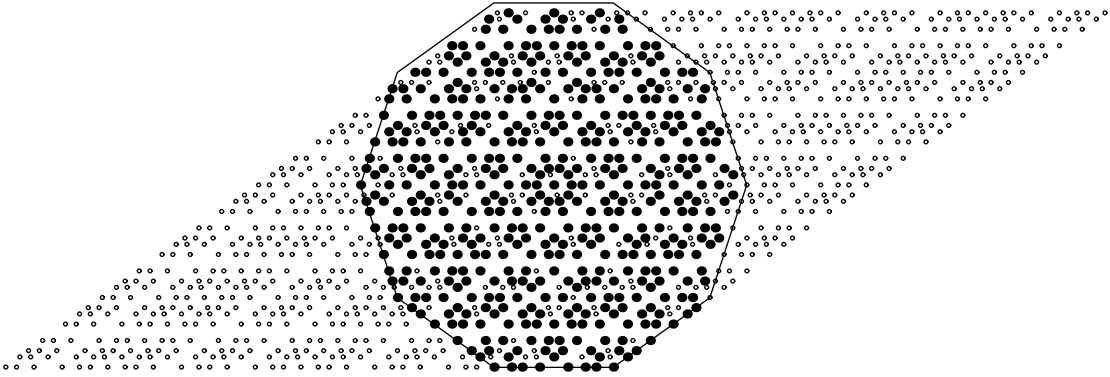}}
\caption{The image $G^{\star}$ of the grid $G$ from Figure~\ref{fig:grid1} under
  the star map together with $S^{\star}$ and the corresponding decagonal window
  already shown in Figure~\ref{fig:slices2}. Note that the star map naturally
  extends to $\Q(\tau)^3$ which contains the grid $G$. Although the
  star map may change the directions and the relative position of
  parallel lines,
  the marginal sums are being preserved. In fact, $G^{\star}$ is the
  grid generated by $S^{\star}$ with respect to the directions that have
  slopes $0$ and
  $\operatorname{tan}(2\pi/10)$, respectively. The reconstruction task is to find a subset of
  $G^{\star}$ that lies in a translate of $L^{\star}$, fits the marginal sums and lies in a suitable
  translate of the interior of the window.}
\label{fig:grid2}
\end{figure}

In the case of planar lattices and {\em two}\/ lattice directions, the above
{\em reconstruction problem}\/ clearly is
intimately related with the problem of reconstructing $0$-$1$-matrices
from their row and column sums. Though, for {\em aperiodic}\/ model sets the
problem is much more involved due to the complication with the
window. Below, an $L$-direction will be called an {\em $L^{(\tau,0,1)}$-direction}\/ 
if it lies in the hyperplane orthogonal to $(\tau,0,1)$, the latter representing a $5$-fold axis of the icosahedral
symmetry of $L$. Observing first that generic F-type icosahedral model sets can be
sliced orthogonal to $(\tau,0,1)$ into (planar) {\em
  cyclotomic model sets} that are based on the ring of integers in the $5$th
cyclotomic field (\cite[Proposition 3.16]{H2}), the following tractability
result can be easily proved by using similar arguments and the same
methods from algebra, graph theory and computational geometry as
in~\cite{BG2}, where  algorithmic complexities in DT of cyclotomic
model sets were studied; see
 Figures~\ref{fig:slices}-\ref{fig:grid2} for illustrations. 

\begin{theorem}{\rm \cite[Theorem 4.3]{H2}}\label{th2}
When restricted to two $L^{(\tau,0,1)}$-directions and polyhedral
  windows, the problem {\sc Reconstruction} can be solved in
  polynomial time in the real RAM-model of computation.
\end{theorem}

\begin{rem}
For a detailed analysis of
the complexities of the reconstruction 
problem in the case of B-type icosahedral model sets, we refer the reader to~\cite[Chapter 3]{H1}. Note 
that even in the case of planar lattices and the Turing machine as
the model of computation the corresponding
reconstruction problem is $\mathbbm{NP}$-hard for three or more
{\em lattice directions} (defined analogous to $\varLambda$-directions); cf.~\cite{GGP}. Therefore, it seems to be rather obvious that
one cannot expect a generalization of Theorem~\ref{th2} to the case of
three or more $L$-directions.
\end{rem}

\section{Uniqueness}\label{sec4}

In general, the above reconstruction problem can possess rather
different solutions. Therefore, one is led to the investigation of the
{\em uniqueness problem}\/ of finding a
 small number of suitably prescribed $L$-directions that
 eliminate these non-uniqueness phenomena. 

\begin{defi}
Let $\mathcal{E}$ be a collection of finite subsets of
$\mathbbm{R}^{3}$ and let $U\subset\mathbb{S}^{2}$ be a
finite set of directions. We say that $\mathcal{E}$ is {\em determined
  by the $X$-rays in the directions of $U$} if different elements of
$\mathcal{E}$ cannot have the same $X$-rays in the directions of $U$.
\end{defi}

Below, a finite subset $C$ of a Delone set $\varLambda \subset \R^3$ is called a {\em convex subset of}\/ $\varLambda$ if  it satisfies the equation  
$C =
\operatorname{conv}(C)\cap \varLambda$; cf.~\cite{GG} for the lattice
case. The set of all convex subsets of
$\varLambda$ will be denoted by $\mathcal{C}(\varLambda)$. Using the
slicing of F-type icosahedral model sets into certain 
cyclotomic model sets again, one obtains the following fundamental result.

\begin{theorem}{\rm \cite[Theorem 5.12]{H2}}\label{main1ico}
\begin{itemize}
\item[{\rm (a)}]
There is a set $U\subset \mathbb{S}^2$ of four pairwise non-parallel 
$L^{(\tau,0,1)}$-directions such that, for all generic F-type icosahedral model sets
$\varLambda$, the set
$\mathcal{C}(\varLambda)$ is determined by the $X$-rays in
the directions of $U$.
\item[{\rm (b)}]
For all generic F-type icosahedral model sets
$\varLambda$ and all sets $U\subset \mathbb{S}^2$  of three or less pairwise
non-parallel $L^{(\tau,0,1)}$-directions, the set
$\mathcal{C}(\varLambda)$ is not determined by the $X$-rays in the
directions of $U$. 
\end{itemize}
 \end{theorem}

In fact, for a generic F-type icosahedral model set
$\varLambda$, the set $\mathcal{C}(\varLambda)$ is determined by the $X$-rays in
the directions of any set $U$ of four pairwise non-parallel 
$L^{(\tau,0,1)}$-directions with the property that there is no {\em
  $U$-polygon}\/ in $\varLambda$; cf.~\cite{H1}. Here, for a set $U$ of $L^{(\tau,0,1)}$-directions, a $U$-polygon in $\varLambda$ is a planar
non-degenerate convex
polygon $P$ with
all its vertices in $\varLambda$ such that any line in $\R^3$ that is
parallel to a direction of $U$ and passes through a vertex of $P$ also
meets another vertex of $P$. It turns out that one can choose four
 pairwise non-parallel $L^{(\tau,0,1)}$-directions which
provide uniqueness and yield dense lines in F-type icosahedral model
 sets; cf.~\cite[Example 5.13 and Remark 5.14]{H2} for examples and
 details. Moreover, there are even four pairwise non-parallel
 $L^{(\tau,0,1)}$-directions which
provide uniqueness, 
 yield dense lines in F-type icosahedral model and have the property
 that, in an
 approximative sense, for any fixed window $W\subset\R^3$ whose boundary $\partial W$ has Lebesgue
measure $0$ in $\R^3$, the set $\cup_{\varLambda\in
   \mathcal{I}^{\rm F}_{g}(W)}\mathcal{C}(\varLambda)$
 is determined by the corresponding $X$-rays; cf.~\cite{H2} for details. This
 setting, 
 which we already used in the definition of the
 reconstruction problem, is highly relevant in practice
 of quantitative HRTEM because it reflects
 the fact that, due to the icosahedral symmetry of genuine
 F-type icosahedral quasicrystals, the determination of the rotational orientation of a
quasicrystalline probe in an electron microscope can rather easily be achieved in the diffraction mode. Though, in general, the $X$-ray images taken
in the high-resolution mode do not allow us to locate the examined sets. Therefore, as already explained in~\cite{BG2,H2}, in order to
prove practically relevant and rigorous results, one has to deal
with the whole local indistinguishability class of a regular, generic
F-type icosahedral model set
$\varLambda$, rather than dealing with a single fixed one.   

\section{Outlook}
Although the presented results answer some of the basic problems of
DT of F-type icosahedral model sets, there is still a lot to
do to create a tool that is as satisfactory for the
application in materials science as is CT in its
medical or other applications. Foremost, this is due to the fact that there is always some noise involved when
physical measurements are taken, whereas the results in this text can
only be applied when exact data is given. Therefore, it is necessary to study stability
and instability results in DT of F-type icosahedral model sets in the
future~\cite{AG}.

\section*{Acknowledgements}
It is a pleasure to thank Michael Baake, Uwe Grimm, Peter Gritzmann, Barbara Langfeld and Reinhard L\"uck for valuable discussions and
suggestions.


\begin{thebibliography}{99}

\bibitem{AG} Alpers, A.; Gritzmann, P.: On stability, error
  correction, and noise compensation in discrete tomography. SIAM
  J. Discrete Math. \textbf{20} (2006) 227--239.

\bibitem{baake} Baake, M.: Solution of the coincidence problem in
     dimensions $d\leq 4$. In: R. V. Moody (Ed.): The Mathematics
   of Long-Range Aperiodic Order. NATO-ASI Series C {\bf 489}, Kluwer,
 Dordrecht (1997), pp. 9--44.; revised version \texttt{arXiv:math/0605222v1 [math.MG]}

\bibitem{B} Baake, M.: A guide to mathematical quasicrystals.
 In: Suck, J.-B.; Schreiber, M.; H\"aussler,
  P. (Eds.): Quasicrystals. An Introduction to Structure, Physical Properties,
    and Applications. Springer, Berlin (2002), pp. 17--48. \texttt{arXiv:math-ph/9901014v1}

\bibitem{BG2} Baake, M.; Gritzmann, P.; Huck, C.; Langfeld, B.; Lord, K.: Discrete tomography of planar model sets.
  Acta Cryst. A{\bfseries 62} (2006) 419--433; \texttt{arXiv:math/0609393v1 [math.MG]} 

\bibitem{BH} Baake, M.; Huck, C.: Discrete tomography of Penrose
    model sets. Philos. Mag. {\bf 87} (2007) 2839--2846; \texttt{arXiv:math-ph/0610056v1}

\bibitem{BM} Baake, M.; Moody, R. V. (Eds.): Directions in Mathematical Quasicrystals. CRM Monograph Series, vol. {\bf 13}, AMS, Providence, RI (2000).

\bibitem{BPR} Baake, M.; Pleasants, P. A. B.; Rehmann, U.: Coincidence
  site modules in $3$-space. Discr. Comput. Geom. {\bf 38} (2006) 111--138; \texttt{arXiv:math/0609793v1 [math.MG]}

\bibitem{HG} de Boissieu, M.; Guyot, P.; Audier, M.: Quasicrystals:
  quasicrystalline order, atomic structure and phase transitions. In:
  Hippert, F.; Gratias, D. (Eds.): Lectures on Quasicrystals. EDP
  Sciences, Les Ulis (1994), pp. 1--152. 

\bibitem{GG} Gardner, R. J.; Gritzmann, P.: Discrete tomography: determination of finite sets by X-rays. Trans. Amer. Math. Soc. {\bf 349} (1997) 2271--2295.

\bibitem{GGP} 
  Gardner, R. J.,  Gritzmann, P., Prangenberg, D.: 
  On the computational complexity of
  reconstructing lattice sets from their X-rays. 
  Discrete Math. \textbf{202} (1999) 45--71.

\bibitem{Gr}
Gritzmann, P.: On the reconstruction of finite lattice sets
from their X-rays. In: E. Ahronovitz; C. Fiorio (Eds.): Lecture Notes on Computer Science, pp. 19--32. Springer, London (1997).

\bibitem{HK} Herman, G. T.; Kuba, A. (Eds.):
 Discrete Tomography: Foundations, Algorithms, and Applications.
Birkh\"auser, Boston (1999).

\bibitem{H1} Huck, C.: Discrete Tomography of Delone Sets with Long-Range Order. PhD Thesis (Universit\"at Bielefeld), Logos Verlag, Berlin (2007).

\bibitem{H2} Huck, C.: Discrete tomography
  of icosahedral model sets. Submitted. \texttt{arXiv:0705.3005v2 [math.MG]}

\bibitem{ks} Kisielowski, C.; Schwander,  P.; Baumann, F. H.; Seibt,  M.; Kim, Y.; Ourmazd,  A.: An approach to quantitative high-resolution transmission electron microscopy of crystalline materials. Ultramicroscopy \textbf{58} (1995) 131--155.

\bibitem{Moody} Moody, R. V.: Model sets: a survey. In: Axel, F.;
  D\'{e}noyer, F.; Gazeau, J.-P. (Eds.): From
    Quasicrystals  to More Complex Systems. EDP Sciences, Les Ulis, and Springer, Berlin (2000), pp. 145--166. \texttt{arXiv:math/0002020v1 [math.MG]} 

\bibitem{sk} Schwander, P.; Kisielowski,  C.; Seibt,  M.; Baumann,  F. H.; Kim,  Y.; Ourmazd,  A.: Mapping projected potential, interfacial roughness, and composition in general crystalline solids by quantitative transmission electron microscopy. Phys. Rev. Lett. \textbf{71} (1993) 4150--4153.



\end{thebibliography}
\end{document}